\documentstyle[epsfig,newpasp,twoside]{article}
\def\etal{{\em et al.}            }
\def\be{\begin{equation}}
\def\ee{\end{equation}}
\def\bea{\begin{eqnarray}}
\def\eea{\end{eqnarray}}

\def\etal{{\em et al.}            }

\def\kms{{\rm \,km\,s}^{-1}}

\def\hMpc{\,h^{-1}{\rm Mpc}}

\def\dg{^{\circ}}

\def\Jy{{\rm \,Jy}}

\def\spose#1{\hbox to 0pt{#1\hss}}
\def\simlt{\mathrel{\spose{\lower 3pt\hbox{$\mathchar"218$}}
     \raise 2.0pt\hbox{$\mathchar"13C$}}}
\def\simgt{\mathrel{\spose{\lower 3pt\hbox{$\mathchar"218$}}
     \raise 2.0pt\hbox{$\mathchar"13E$}}}
\def\({\left(}
\def\){\right)}
\def\[{\left[}
\def\]{\right]}
\def\<{\left\langle}
\def\>{\right\rangle}

\def\ApJ{{\em ApJ~}}

\def\IAU130{in {\em Large Scale Structures of the Universe}, IAU Symposium 130~}
\def\MN{{\em MNRAS~}}

\def\edcomment#1{\iffalse\marginpar{\raggedright\sl#1\/}\else\relax\fi}
\reversemarginpar

\begin{document}
\title{The IRAS view of the Local Universe}
\author{Will Saunders$^1$, Kenton D'Mellow$^1$, Helen Valentine$^1$, Brent Tully$^2$}
\affil{$^1$University of Edinburgh, UK. $^2$University of Hawaii, USA.}
\author{Esperanza Carrasco$^1$, Bahram Mobasher$^2$, Steve Maddox$^3$, George Hau$^4$}
\affil{$^1$INAOE, Mexico. $^2$STSCI, USA. $^3$University of Nottingham, UK. $^4$Universidad Catholica, Santiago, Chile.}
\author{Will Sutherland$^1$, Dave Clements$^2$, Lister Staveley-Smith$^3$}
\affil{$^1$ATC, ROE, UK. $^2$ University of Cardiff, UK. $^3$CSIRO, Australia.}

\begin{abstract}

We summarise results for the predicted density and peculiar velocity
fields from the PSCz survey, consisting of redshifts for 15,000 IRAS
galaxies covering 84\% of the sky to a depth of $25,000\kms$. We have
used a generalisation of the Path Interchange Zeldovich Approximation
technique to determine the velocity field; the most remarkable feature
being a coherent large-scale flow along the
baseline connecting the Local Supercluster, Centaurus
and the Shapley Concentration. Comparison of the predicted and
observed bulk flows gives a value of $\beta = 0.50\pm0.1$.

We re-examine the PSCz dipole, with improved redshift completeness at low-latitudes and using PIZA to estimate real-space distances. We find the dipole to be stable between $80$ and $180 \hMpc$, although there appears to be a significant contribution to the dipole around $200 \hMpc$. The overall misalignment with the CMB dipole remains at $20\dg$. The implied value of $\beta$ depends on the exact treatment; we derive values in the range $\beta = 0.40-0.55$ with statistical error $\pm0.1$.

We also present the density field and a preliminary dipole analysis
from the almost completed Behind The Plane survey, which extends the
PSCz to cover 93\% of the sky including the proposed core of the Great
Attractor.  We find a density peak at $(325,-5,3500\kms)$, about half
as massive as Centaurus or Pavo-Indus, and forming part of a
continuous filament linking them across the Plane. We also find
evidence for a much larger `Greater Attractor' directly behind the GA, at a
distance of $125 \hMpc$, and more massive than the Shapley
concentration.
 
At large distances the dipole direction is in much improved
agreement with the CMB: at $250-300\hMpc$, the misalignment is only
$5-10\dg$, and this is consistent with the shot noise errors. The $J_3$-weighted dipole, which in principal is expected to agree better with
the CMB, gives a misalignment of $13\dg$. The dipole direction is somewhat
dependent on the corrections made for the redshift incompleteness, and
may also be affected by unquantified incompleteness in the BTP survey
close to the Galactic Centre. The dipole amplitude implies a value for
the parameter $\beta$ of $0.44\pm0.1$. Values much larger than
$\beta=0.5$, while not formally ruled out, cause very unlikely-looking
rocket effects.

\end{abstract}

\section{Introduction}
\vspace{-5pt}

Redshift surveys of galaxies can be used to infer the dynamics of the Local Universe. Comparison with observed peculiar velocities allows determination of the mass associated with galaxies (usually parameterised by the the quantity $\beta \simeq \Omega^0.6/b$), and also reconstruction of the initial conditions. Because gravity is extensive, accurate predictions formally require uniform surveys with infinite depth and complete sky coverage; failing that, we need a depth much larger than the region of interest, and with coverage gaps small enough to interpolate over. IRAS surveyed 96\% of the sky with negligible extinction, tracing the large-scale structure to a depth of at least $20,000 \kms$. For these reasons, IRAS-based surveys have dominated this field for the last decade. The IRAS dipole has remained astonishingly stable in direction over this period, despite the ever-increasing depth and sky coverage of the follow-up redshift surveys. However, the amplitude has increased continuously, with extra contributions uncovered at larger distances and lower latitudes.

The Point Source Calatog Redshift Survey (PSCz) consists of essentially all galaxies in the PSC, down to its nominal flux limit of $0.6\Jy$, across the 84\% of the sky where extinction was estimated to be low enough ($A_B<2^m$) to allow reasonably complete optical identifications and spectroscopy with 2-metre class telescopes. The PSCz data is now available from http://www-astro.physics.ox.ac.uk/$\sim$wjs/pscz.html; also there available is the full, labelled 3D density distribution, using the Fourier interpolation method presented in Saunders and Ballinger (2000). The catalogue and main science papers to date are summarised in Saunders \etal (2000a).

The PSC itself is useable much deeper into the Plane. The Behind The Plane Survey extends the PSCz to cover the entire 93\% of the sky with useable PSC data, and is now almost completed. The remaining size of the Zone of Avoidance is just 4\% of the sky, and this is hopefully small enough to interpolate over with confidence. The sky distribution for both surveys is shown in Figure 1.

\begin{figure}
\centerline{\epsfig{figure=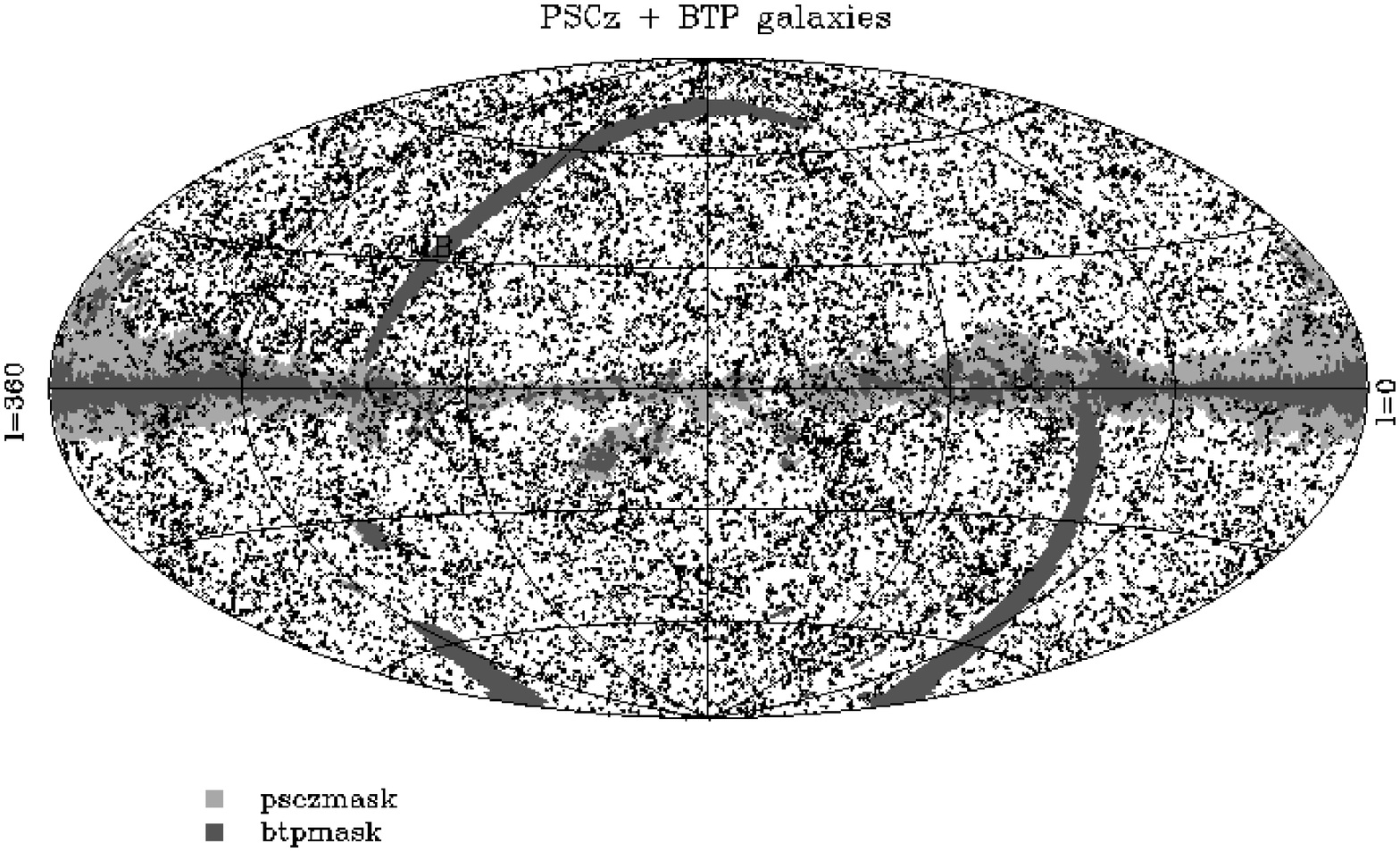,width=8cm,angle=-90}}
Figure 1. PSCz and BTP galaxy catalogues and masks in galactic coordinates.
\vspace{-5pt}
\end{figure}

\section{The PSCz velocity field and the Path Interchange Zeldovich Approximation}
\vspace{-5pt}

A linear theory reconstruction of the velocity field and comparison with the Mark III dataset was been performed by Branchini \etal (1999). However, the heavy smoothing needed to make linear theory applicable seriously compromises the detail with which such comparisons can be done. 

The PIZA algorithm, originally presented by Croft and Gazta\~{n}aga
(1997), is a quasi-linear method for reconstructing the velocity field
from galaxy surveys, allowing the full resolution of the data to be
kept. The basic idea is wonderful and very simple: to Zeldovich order,
velocities and trajectories depend in a fixed way on the scale factor;
so minimising the action is equivalent to minimising the rms
trajectories of the galaxies. Given a set of uniform initial positions
and the current galaxy positions, the least-action solution is found
by setting up arbitrary trajectories and swapping pairs to reduce the
action. The formalism can be readily adapted to minimise (for a given
$\beta$) the real-space trajectories from redshift-space
information. With more complexity, the changes in selection function
for individual galaxies caused by peculiar velocities can also be
included, though we have found doing this leads to instabilities
without any improvement in accuracy, and this is not done here. This
and edge effects lead to a systematic velocity underestimation of
$\sim20\%$ in comparison with simulations. Further details are
presented in Valentine, Saunders and Taylor (2000). The resulting
velocity field in the Supergalactic Plane is shown in Figure 2. The
flow field is dominated by infall into the Supergalactic Plane, and
coherent flow along it involving the Local Supercluster, Centaurus and
the Shapley Concentration.

\section{Comparison with observed peculiar velocities}

The velocity field predicted from PIZA can be compared with observations. The derived trajectory of the Local Group is robust, with weak dependence on how the mask is filled or the input $\beta$. Comparison of this trajectory with the CMB dipole gives $\beta=0.53\pm0.1$. A comparison of the bulk flow as a function of radius with the MarkIII dataset implies a value for $\beta$ of $0.5\pm 0.15$.

We have also undertaken a new comparison with the SFI dataset of peculiar velocities of Spiral Galaxies (Giovanelli \etal 1998). We have calculated the average peculiar velocity with respect to the Local Group, of galaxies in redshift slices, so as to be as closely as possible comparable to Giovanelli \etal. We find reasonable agreement in amplitude, though there are significant directional differences, for $\beta=0.55\pm 0.1$ (Table 1). We agree with Giovanelli \etal that most of the dipole is generated within $5000 \kms$. However, we do find substantial contribution to the dipole beyond this distance (see below), even though our derived velocity dipole at this distance is in excellent agreement with the CMB. The sum derived by Giovanelli \etal is a {\em galaxy}-weighted vector sum, rather than volume-weighted. Therefore, a good agreement between velocity vector and CMB dipole does not necessarily imply that a cosmic rest frame has been reached.

\begin{table}
\begin{tabular}{rcccc}
Shell & $V_{SFI}$ & $(l,b)_{SFI}$ & $V_{PIZA}$ & $(l,b)_{PIZA}$ \\
0-2000    & $270\pm80$ & $(245,49)\pm19$ & 338 & (237,41) \\
1500-3500 & $410\pm69$ & $(255,21)\pm12$ & 404 & (265,51) \\
2500-4500 & $620\pm76$ & $(255,15)\pm11$ & 471 & (271,45) \\
3500-5500 & $585\pm92$ & $(265,19)\pm13$ & 550 & (285,36) \\
4500-6500 & $544\pm98$ & $(270,16)\pm15$ & 508 & (284,31) \\
\end{tabular}

Table 1. SFI vs PSCz peculiar velocities with respect to shells of galaxies. PSCz reconstruction assumes $\beta=0.5$.
\end{table}

\begin{figure}
\centerline{\epsfig{figure=saunders-a2.eps,width=17cm}}
Figure 2. Reconstructed velocity field in the Supergalactic Plane. The Great Attractor region is at 10 o'clock.

\vspace{-15pt}
\end{figure}

\section{PSCz dipole}

In the course of the BTP survey, we also took several hundred new reshifts for low-latitude PSCz galaxies, giving an overall redshift incompleteness for bright galaxies of 70 galaxies (0.5\%). A further 230 galaxies (1.5\%) without redshifts are assumed from their optical and/or infrared appearance to be beyond $30,000\kms$. In the original, publicly available survey, there was known incompleteness at lower redshifts towards the Galactic anti-centre as discussed by Saunders \etal (2000a) and whose effect on the dipole is discussed in Rowan-Robinson \etal (2000). It is appropriate then to recalculate the dipole analysed by Rowan-Robinson \etal. We have corrected for peculiar velocities using PIZA as above, with $\beta=0.5$ for self-consistency between gravity and velocity dipoles. The mask is filled in via the Fourier interpolation method of Saunders and Ballinger (2000). The revised dipole is presented in Figure 3. The dipole is remarkably stable between $125-180\hMpc$, However, it shows a significant shift at $180-220 \hMpc$ in the X and Y components. This shift is seen in the analysis of Rowan-Robinson \etal, and is too large to be attributed to the small remaining redshift incompleteness. It remains when the interpolated galaxies in the Plane are replaced by randomly distributed galaxies at the mean density, so it is not due to the interpolation either. It does not appear to be associated with any single particular structure.

In linear theory, the contribution to the velocity dipole per galaxy is 

\be
{\bf V(r)} = {\beta H_0\over{4 \pi}} {{\bf r}\over{r^3 \psi({\bf r})}}
\ee

\noindent where $\psi$ is the selection function. In Figure 4, we plot the amplitude of this velocity as a function of distance for the PSCz. The dipole contribution per galaxy is remarkably constant in the range $20-100\hMpc$. Nearby, the contribution is large because of the $1/r^2$ term, but at least the numbers of galaxies involved are small and the structures they reside in are real and well-sampled. At large distances, the contribution blows up because $\psi$ falls off much faster than $1/r^2$, and the dipole becomes increasingly dominated by shot noise (e.g. Taylor and Valentine 1999). Strauss \etal (1992) derived an optimal weighting scheme, in the sense of producing the minimum variance ensemble difference between true and estimated gravity vectors. At large distances, this amounts to giving each galaxy a `$J_3$ weighting' i.e.

\be
{\bf V} = {\beta H_0\over{4 \pi}} \sum_{gals}{{J_3(r) \psi({\bf r})} \over{1+ {J_3(r) \psi({\bf r})}}} {{\bf r}\over{r^3 \psi({r})}}
\ee

Where $ J_3(r) = 4 \pi \int_0^r r'^2 \xi(r') dr'$ is the integral over the correlation function. $J_3(r)$ is not well-determined on large scales, but the precise value used is not critical as any error only means the weighting is quite optimal, while leaving the dipole estimate unbiased in an ensemble sense. Based on the PSCz power spectrum (Sutherland \etal 1999), we have simply assigned $J_3(r)$ a constant value $J_3=20,000 \hMpc$ for all $r$. The resulting velocity dipole per galaxy is shown in Figure 4; there is now a sharp attenuation above $200\hMpc$.

Based on Lineweaver \etal (1996) and Yahil, Sandage and Tamman (1977), we have assumed a value for the LG motion with respect to the CMB of $(276,30,627 \kms)$. The misalignment of CMB and IRAS dipoles remains $15-25\dg$, depending on chosen upper cutoff radius for the gravity sum, precise treatment of the mask etc. The inferred value of $\beta$ is $\beta=0.54\pm0.10$ from the overall amplitude; however, the Z-component, which is very stable and also less sensitive to the inferred structure behind the Plane, gives $\beta=0.40\pm0.07$.

The dipole and peculiar velocity results are largely independent, since the latter is sensitive only to tidal effects from large scales. The completely independent analysis of Tadros \etal (1999), using the statistical distortion of the clustering in redshift space, also gives a consistent, low-$\beta$ result $\beta=0.47\pm0.16$.

\begin{figure}
\centerline{\epsfig{figure=saunders-a3.eps,width=13.4cm}}
Figure 3(a). X,Y,Z and total velocity components of the real-space PSCz dipole, with redshifts corrected for peculiar motions using the PIZA reconstructions, with an assumed value of $\beta=0.5$. The dashed lines show the $J_3$-weighted components.\\
Figure 3(b). Angular components in Galactic coordinates; showing direction at $100,200$ and $300\hMpc$ (filled dots) with their one-sigma shot noise (dotted circles), also the $J_3$-weighted direction (cross) and associated one-sigma shot noise (dashed circle), and the CMB dipole (star) and circles showing $5\dg$ and $10\dg$ deviation. 
\end{figure}

\begin{figure}
\centerline{\epsfig{figure=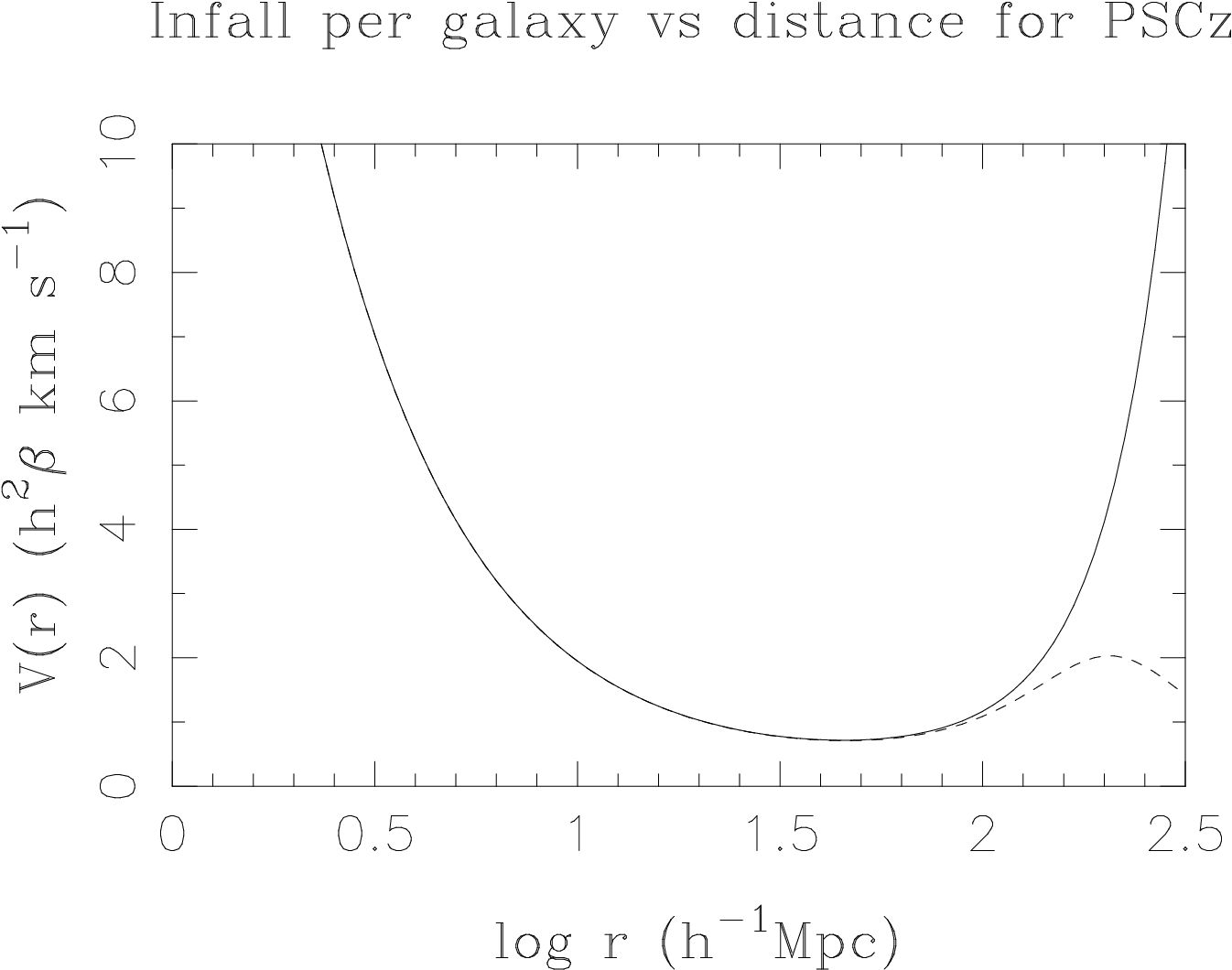,width=6cm,angle=-90}}
Figure 4. Linear theory velocity generated at the Local Group for a galaxy in the PSC for a galaxy of given distance. The dashed line shows the effect of the $J_3$-weighting.
\end{figure}

\section{The Behind The Plane survey}

The 84\% sky coverage of the PSCz survey is effectively limited by the need to get, for every galaxy, an optical identification from sky survey plates. The fractional incompleteness in sky coverage translates directly into a much larger uncertainty in predicting the gravity dipole on the Local Group: suppose the fractional masked area is $f$, and also that this is large and compact enough that we have essentially no information about the distribution behind the mask. Then we can think of the sky as consisting of $1/f$ independent contributions to the dipole, one of which is unknown, giving a fractional uncertainty in the measured dipole vector of $\sqrt{f/(1-f)}$. Thus even 10\% missing sky coverage can lead to a error of 1/3 in the dipole. The bulk of the dipole is generated around $40\hMpc$, at which distance the correlation length $r_0$ subtends just $5\dg$, so gaps much larger than this are too large to interpolate usefully and lead to large errors in the dipole.

The IRAS PSC data itself is reliable to much lower latitudes, although genuine galaxies are outnumbered by Galactic sources with similar IRAS properties. Previous attempts to go further into the Plane have either been restricted to the Arecibo declination range, or have relied on optical identifications from Sky Survey Plates. Because the extinction may be several magnitudes or more, they have inevitably suffered from progressive and unquantifiable incompleteness as a function of latitude. Hence in 1994 we embarked on a program, parallel with the PSCz survey, to systematically identify and get redshifts for low latitude IRAS galaxies wherever the PSC data allowed - essentially limited only by IRAS coverage and confusion, to 93\% of the sky. A detailed description is given in the associated poster paper (Saunders \etal 2000b). The identification program is complete; and we have found over 200 galaxies which were not included in the 1.2Jy survey but should have been. The BTP sky distribution is shown along with the PSCz in Figure 1. At low latitudes towards the Galactic centre, the PSC is not complete to our nominal flux limit of $0.6\Jy$, and the density of identified BTP galaxies reflect this. This incompleteness is interpreted, quantified and corrected for in the poster paper. Notwithstanding this, the huge overdensity of galaxies in the Great Attractor region $(325,\pm 5)$ is obvious; as is the Puppis supercluster $(250,0)$ and the extension from Perseus (150,-10) and Pegasus (95,-10) into the Plane.

 The overall surface density of IRAS galaxies in the BTP area, when corrected for known incompleteness, is 78\% of the PSCz average. Unquantified incompletenesses could conceivably push this up to the average value, but clearly we can rule out large overall overdensity behind the Plane. We have used the Fourier interpolation technique to find the projected surface density of the PSCz+BTP galaxy distribution; this is shown in Figure 5. In addition to the above features, we see the Ophiucus Spercluster clearly at (0,10) and the local void behind the Galactic Centre.

\begin{figure}
\vspace{-10pt}
\centerline{\epsfig{figure=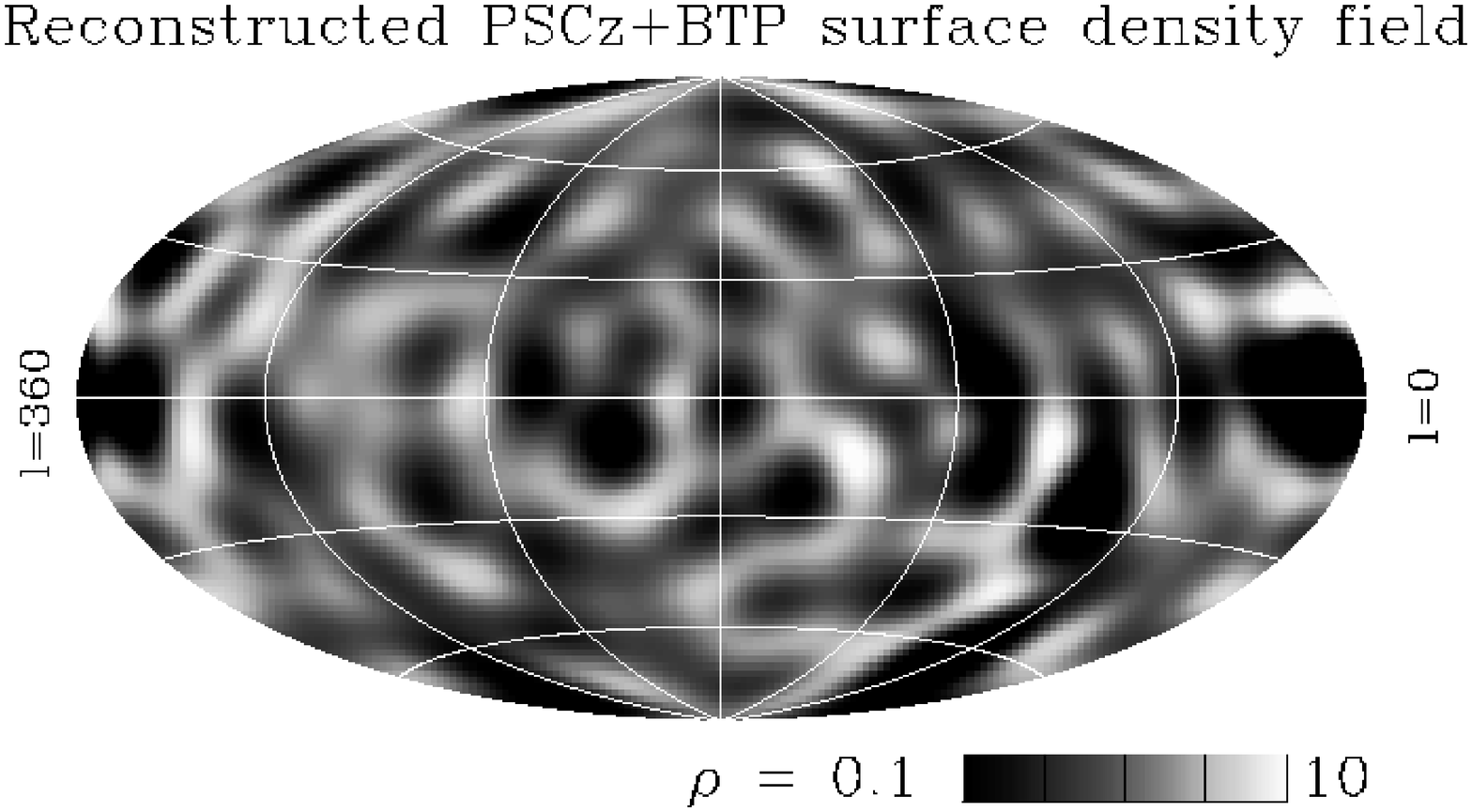,width=8cm,angle=-90}}
Figure 5. Surface density of PSCz+BTP galaxies in galactic coordinates.
\end{figure}
\begin{figure}
\centerline{\epsfig{figure=saunders-a6.eps,width=9cm,angle=-90}}
Figure 6. Weighted N(z) distribution for the BTP survey. Open circles show the effect of assigning a nearest-neighbour redshift to galaxies without measured redshifts themselves. The line is the prediction from the high-latitude PSCz.
\end{figure}

\section{3D distribution}

The spectroscopy is now completed in the south, and useably completed in the north. In addition to BTP galaxies, we also systematicall pursued faint, low-latitude PSCz galaxies where there was any likelihood that these were within $30,000\kms$. There are c.370 very faint galaxies in the combined PSCz+BTP with unknown redshift, but these are likely beyond $30,000\kms$. There are a further 240 for which we would like a redshift but do not currently have one, though for half of these we have unreduced data. As a temporary fix, for these latter 240 galaxies we have simply assigned the redshift of a nearby galaxy with redshift. The N(z) distribution, both with and without this fix, is shown in Figure 6, together with the prediction from the PSCz survey. There is some evidence that we are missing galaxies at most or all redshifts, but also that there is no cutoff in the redshift completeness with distance till at least $25,000 \kms$ - much better than our original target of $15,000 \kms$.

We can then make a 3-dimensional map of the galaxy distribution. The full, animated 3D map is available from http://www.roe.ac.uk/willwww. Figure 7 shows the peak we find in the GA region at $(325,-5,3500\kms)$. The overdensity is quite modest compared with other superclusters in the region, and the peak overdensity is closer than the velocity of the Norma cluster at $4844\kms$ (Woudt \etal 2000). However, we find the GA to be very extended in redshift space, from $3000-5500\kms$. More surprising is Figure 8, showing an apparent `Greater Attractor', directly behind the GA at $(326,-3,12500\kms)$. Because of the incompleteness weighting, and because the overdensity is right on the edge on the mask, the finding represents only a dozen measured redshifts and is at this stage preliminary.

\begin{figure}
\centerline{\epsfig{figure=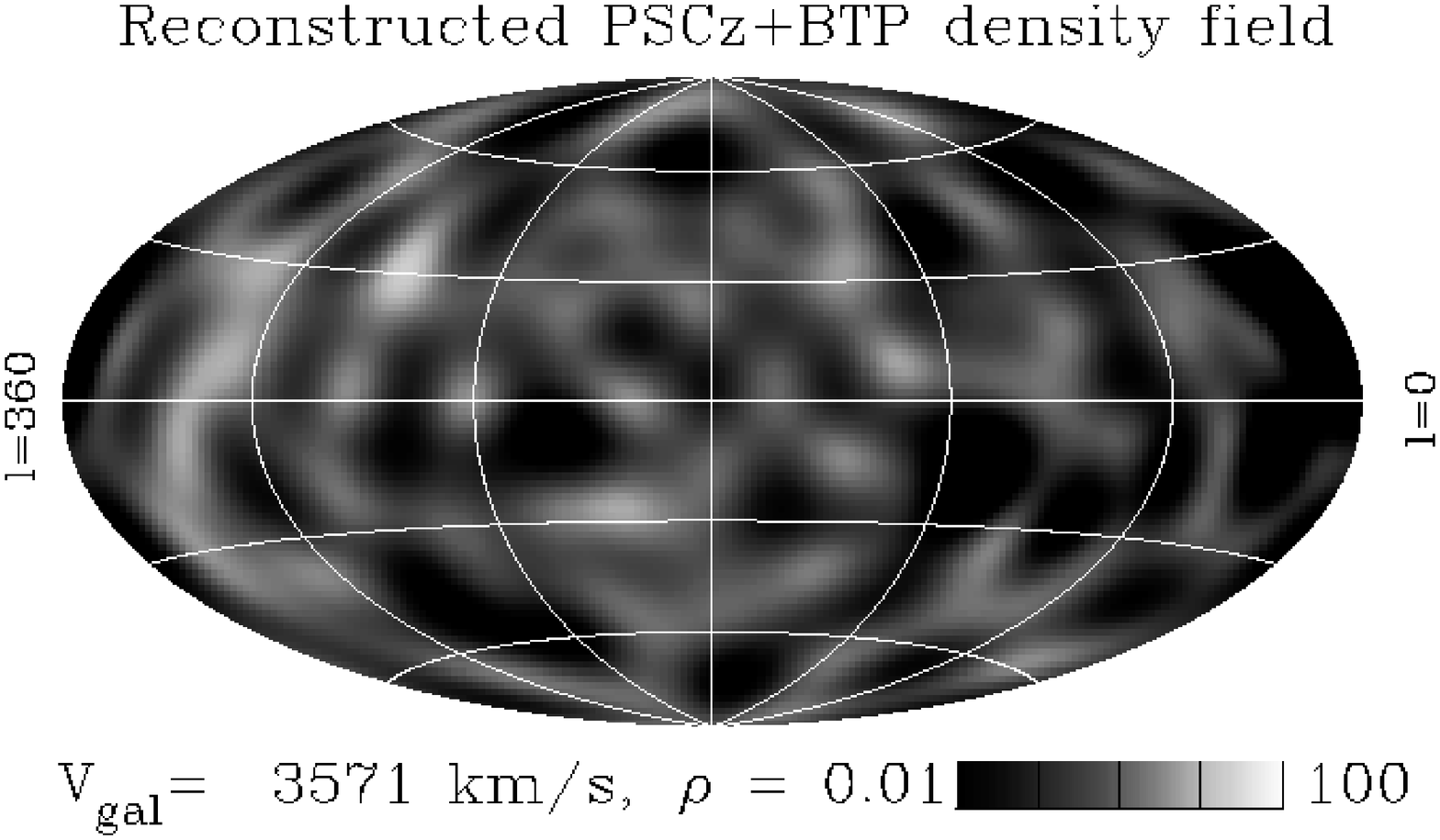,width=8cm,angle=-90}}
Figure 7. Galaxy density field at a redshift of $3571 \kms$ in galactic coordinates, showing the peak of the Great Attractor at (325,-5).
\vspace{-15pt}
\end{figure}
\begin{figure}
\centerline{\epsfig{figure=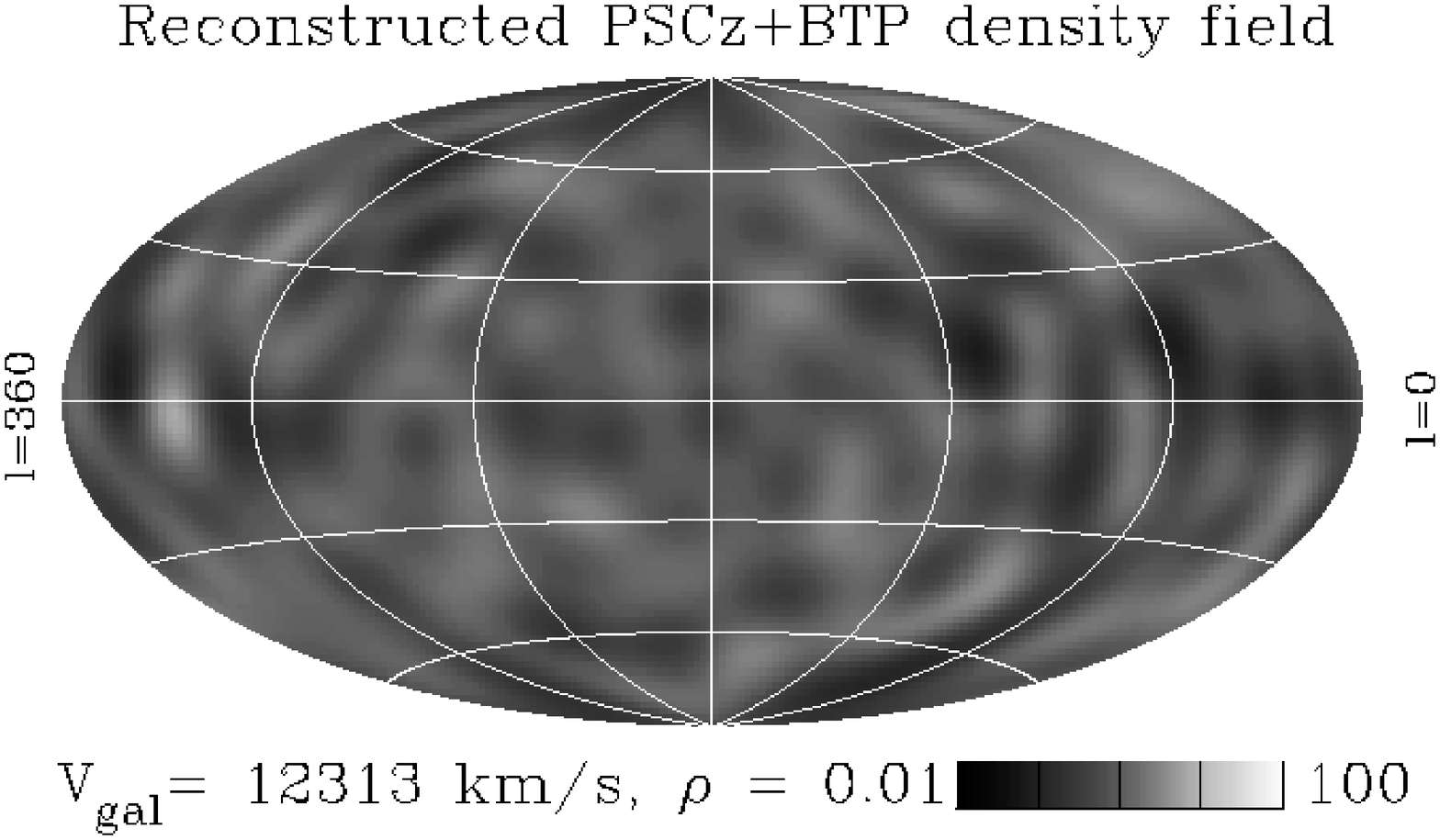,width=8cm,angle=-90}}
Figure 8. Galaxy density field at a redshift of $12313 \kms$ in galactic coordinates, showing the `Greater Attractor' at (326,-3).
\vspace{-15pt}
\end{figure}

\section{The PSCz+BTP Dipole}

The combined PSCz+BTP sample is now complete enough to attempt a revised dipole. One of the main motivations for this is that current low $\Omega$ models predict small misalignments between gravity and velocity vectors, especially as the Hubble flow in the immediate vicinity of the Local Group (where non-linear effects are most likely to arise) is so cold: a misalignment as large as the $20-25\dg$ between IRAS and CMB dipoles would present problems for current models. As above, we have assigned a nearest neighbour redshift to 240 galaxies. We have used our interpolated 3D maps to generate random `galaxies' to add to the catalogue when determining the dipole. However, because the missing region is so much smaller, the results are now much less sensitive to this interpolation.

To get real-space positions, we used the PIZA reconstruction for the PSCz galaxies as above, with an assumed $\beta=0.5$. For the BTP sources we just used the redshift-space position. The resulting dipole is shown in Figure 9. Overall, the dipole looks much the same as for the PSCz alone. However, the misalignment with the CMB is now very much smaller, being $10\dg$ at $250\hMpc$ and $5\dg$ at $300\hMpc$, with the shot noise error being $10\dg$. The misalignment is $13\dg$ for the $J_3$-weighted sum. The directional agreement also appears to be robust: if the galaxies with unknown redshift are dropped altogether, the dipole only changes by $5\dg$ in direction and 7\% in amplitude, while if random sources are put uniformly in the mask at the average density instead of interpolated, the dipole only alters by a few degrees in direction and 10\% in amplitude.

The shift at $180-220 \hMpc$ seen in the PSCz remains. At this stage, it remains possible that variable redshift completeness is responsible. Also, at this distance, very slight systematic effects in the PSC itself could have large effects on the dipole, but these would not be expected to be restricted to a particular velocity range as seen in Figure 8. Beyond that, there is a drift in the X component, but this too is sensitive to variable redshift incompleteness around the Galactic Plane and is not secure at this stage.

Peacock (1992) noted that apparent convergeance in the dipole may be illusory. The apparent convergeance of the IRAS dipole between $80-180 \hMpc$, in a direction only $20\dg$ away from the CMB, was seen in the QDOT, 1.2Jy and PSCz surveys, and was in each case interpreted as the `true' value for the determination of $\beta$, with no significant contribution from larger distances. We now see that there may indeed be a significant contribution from beyond $180\hMpc$. Of course the same argument can in principal be applied this work, in that we have not ruled out contributions from beyond $300\hMpc$. However, we can hope that the excellent directional agreement we find, combined with the implausible size of structures required, make this unlikely.

\begin{figure}
\centerline{\epsfig{figure=saunders-a9.eps,width=13.4cm}}

Figure 9(a). X,Y,Z and total velocity components of the real-space PSCz+BTP dipole, with PSCz redshifts corrected for peculiar motions using the PIZA reconstructions, with an assumed value of $\beta=0.5$. The dashed lines show the $J_3$-weighted components.
Figure 9(b). Angular components in Galactic coordinates; showing direction at $100,200$ and $300\hMpc$ (filled dots) with their one-sigma shot noise (dotted circles), also the $J_3$-weighted direction (cross) and associated one-sigma shot noise (dashed circle), and the CMB dipole (star) and circles showing $5\dg$ and $10\dg$ deviation. 
\vspace{-15pt}
\end{figure}

\end{document}